\documentstyle[11pt,a4,epsfig]{article}

\makeatletter
\@addtoreset{equation}{section}
\makeatother

\newcommand{\comment}[1]{{}}

\newcommand{\beq}{\begin{equation}}
\newcommand{\eeq}{\end{equation}}
\newcommand{\ba}[1]{\begin{array}{#1}}
\newcommand{\ea}{\end{array}}
\newcommand{\bea}{\begin{eqnarray}}
\newcommand{\eea}{\end{eqnarray}}

\newcommand{\sz}{\scriptsize}

\newcommand{\bm}[1]{\mbox{\boldmath $#1$}}

\newcommand{\jump}[1]{$^{\mbox{\sz {#1}}}$\/}

\title{
{\bf
On the determination of probability density functions by using
Neural Networks
} \\
\vspace{1ex}
}

\date{}

\author{
\large {\sc Llu\'{\i}s Garrido\jump{1,2}\/, Aurelio Juste\jump{2}\/} \\
\\
\normalsize  1) Dept.\ d'Estructura i Constituents de la Mat\`eria, \\
\normalsize  Facultat de F\'{\i}sica, Universitat de Barcelona, \\
\normalsize  Diagonal 647, E-08028 Barcelona, Spain. \\
\normalsize  Phone: +34\,93\,402\,11\,91 \ \ Fax: +34\,93\,402\,11\,98 \\
\normalsize  e-mail: garrido@ecm.ub.es \\
\\
\normalsize  2) Institut de F\'{\i}sica d'Altes Energies, \\
\normalsize  Universitat Aut\`{o}noma de Barcelona, \\
\normalsize  E-08193 Bellaterra (Barcelona), Spain. \\
\normalsize  Phone: +34\,93\,581\,28\,34 \ \ Fax: +34\,93\,581\,19\,38 \\
\normalsize  e-mail: juste@ifae.es \\
}

\begin{document}

\maketitle
\begin{abstract}
It is well known that the output of a Neural Network trained to disentangle between two classes
has a probabilistic interpretation in terms of the a-posteriori Bayesian
probability, provided that a unary representation is taken for the output patterns.
This fact is used to make Neural Networks approximate probability density functions
from examples in an unbinned way, giving a better performace than
``standard binned procedures''. In addition, the mapped p.d.f. has an analytical expression.

\vspace{2ex}
\begin{center} PACS'96: 02.50.Ph, 07.05.Kf, 07.05.Mh \end{center}
\vspace{1ex}
\begin{center} (Submitted to {\sl Comput.\ Phys.\ Commun.}) \end{center}
\end{abstract}

\newpage

\section{Introduction}

Estimating a probability density function (p.d.f.) in a $n$-dimensional space is a
necessity which one may easily encounter in Physics and other fields.
The standard procedure is to bin the space and approximate
the p.d.f. by the ratio between the number of events falling inside
each bin over the total and normalised to the bin volume. 
The fact of binning not only leads to a loss of information 
(which might be important unless the function is smoothly varying inside each
bin) but is intrinsically arbitrary: no strong arguments for a defined binning strategy,
e.g. constant bin size versus constant density per bin, exists.
More sophisticated approaches imply for instance
the definition of an ``intelligent'' binning, with
smaller bins in the regions of rapid function variation.
However, the main drawback still remains:
even for a low number of bins per dimension, large amounts of 
data are necessary since the number of data points needed to fill the bins
with enough statistical significance grows exponentially with the number of variables.
As it will be shown,
Neural Networks (NN) turn out to be useful tools for building up
analytical $n$-dimensional probability density functions in an unbinned way from examples.

This manuscript is organised as follows: in Sect. 2 the proposed method to construct
unbinned p.d.f.s from examples is described. After a brief introduction to the statistical
interpretation of the output of a Neural Network applied to pattern recognition in the
case of only two classes, an expression for the mapped p.d.f. is obtained. Then, a method
to quantify the goodness of the mapped p.d.f. is described.
In order to illustrate the concept, an artificial example is discussed in Sect. 3, whereas Sect. 4
is devoted to the discussion of an example of practical application in High Energy Physics.
Finally, in Sect. 5, the conclusions are given.

\section{Method}

Let us assume that we have a sample of $N$ events distributed among 2 different classes of patterns
(${\cal C}_1$ and ${\cal C}_2$), each event $e$ being characterised by a set of $n$ variables ${\bm x}^{(e)}$.
Each class of patterns has a proportion $\alpha_i$ and is generated by the normalised probability
density function $P_i({\bm x})$, $i=1,2$ (in probability terms, $P_i({\bm x}) = P({\bm x}\mid {\cal C}_i)$ and
$\alpha_i = P({\cal C}_i)$).

By minimising over this sample the quadratic output-error $E$:
\begin{equation}
  E\left [o\right ] = \frac{1}{2N} \sum_{e=1}^N 
\left [o({\bm x}^{(e)})-d({\bm x}^{(e)})\right ]^2.
\end{equation}
\noindent with respect to the unconstrained function $o({\bm x})$, where $d({\bm x})$
takes the value 1 for the events belonging to class ${\cal C}_1$ and 
0 for the events belonging to class ${\cal C}_2$, it can be shown \cite{Garrido,Papoulis,Ruck,Wan} 
that the minimum is achieved when $o({\bm x})$ is the a-posteriori Bayesian
probability to belong to class ${\cal C}_1$:
\begin{equation}
 {\em o}^{(min)}({\bm x}) = {\cal P}({\cal C}_1 \mid {\bm x}).
\end{equation}

The above procedure is usually done by using 
layered feed-forward Neural Networks (see e.g.~\cite{Hertz,Muller} for an introduction).
In this paper we have considered Neural Networks 
with topologies $N_i \times N_{h_1} \times N_{h_2} \times N_o$, where $N_i$ ($N_o = 1$) are the number of input
(ouput) neurons and $N_{h_1}$, $N_{h_2}$ are the number of neurons in two hidden layers.

The input of neuron $i$ in layer $\ell$ is given by,
\begin{equation}
  I_{i}^{\ell} =\left \{
\begin{array}{cl}
x_{i}^{(e)} & \quad \ell=1 \\
\sum_{j} w_{ij}^{\ell} S_{j}^{\ell-1} + B_{i}^{\ell} & \quad \ell=2,3,4 \\
\end{array}
\right.
\end{equation}
\noindent where $x_{i}^{(e)}$ is the set of $n$ variables describing a physical event $e$, the sum is
extended over the neurons of the preceding layer $(\ell-1)$, $S_{j}^{\ell-1}$ is the state of
neuron $j$ at layer $(\ell -1)$
and $B_{i}^{\ell}$ is a bias input to neuron $i$ at layer $\ell$. The state of a neuron is a function
of its input $S_{j}^{\ell} = F(I_{j}^{\ell})$, where $F$ is the neuron response function. In
general the ``sigmoid function'', $F(I_{j}^{\ell}) = 1/(1+e^{-I_{j}^{\ell}})$,
is chosen since it offers a more sensitive modeling of real data than a linear one, being able
to handle existing non-linear correlations. 
However, depending on the particular problem faced, a different
neuron response function may be more convenient. For instance, in the artificial example described below, a
sinusoidal neuron response function, $F(I_{j}^{\ell}) = (1+\sin(I_{j}^{\ell}))/2$, has been adopted.

Back-propagation \cite{Rumelhart1,Rumelhart2,Werbos} is used as the learning algorithm.
Its main objective is to minimise the above quadratic output-error $E$
by adjusting the $w_{ij}$ and $B_{i}$ parameters.

Let us now consider the situation we are concerned in this paper: 
we have a large amount of events (``data'') distributed according
to the p.d.f. ${\cal P}_{data}({\bm x})$, whose analytical expression is unknown and  which we want precisely to approximate.
If a Neural Network is trained to disentangle between those events and other
ones generated according to any kwown p.d.f., ${\cal P}_{ref}({\bm x})$ (not vanishing in
a region where ${\cal P}_{data}({\bm x})$ is non-zero),
the Neural Network output
will approximate, after training, the conditional probability for a given
event to be of the ``data'' type:

\begin{equation}
{\em o}^{(min)}({\bm x}) \simeq {\cal P}(data \mid {\bm x}) \equiv \frac{\alpha_{data} {\cal P}_{data}({\bm x})}
{\alpha_{data} {\cal P}_{data}({\bm x})+\alpha_{ref} {\cal P}_{ref}({\bm x})},
\end{equation}

\noindent where $\alpha_{data}$ and $\alpha_{ref}$ are the proportions
of each class of events used for training, satisfying
$\alpha_{data}+\alpha_{ref} = 1$. 

From the above expression it is straightforward to extract the NN approximation to ${\cal P}_{data}({\bm x})$
as given by:

\begin{equation}
{\cal P}_{data}^{(NN)}({\bm x}) = {\cal P}_{ref}({\bm x}) \frac{\alpha_{ref}}{\alpha_{data}}
\frac{{\em o}^{(min)}({\bm x})}{1-{\em o}^{(min)}({\bm x})}.
\label{pdf}
\end{equation}

As a result, the desired p.d.f. is determined in an unbinned way from examples.
In addition, ${\cal P}_{data}^{(NN)}({\bm x})$ has an analytical expression since 
we indeed have it for ${\cal P}_{ref}({\bm x})$ and $o^{(min)}({\bm x})$ is known once we have
determined the network parameters (weights and bias inputs).

For what the reference p.d.f. is concerned, a good choice would be a p.d.f.
built from the product of normalised good approximations to each 1-dimensional projection of the data p.d.f., thus
making easier the learning of the existing correlations in the $n$-dimensional space.
Since ${\cal P}_{ref}({\bm x})$ is a normalised p.d.f. by construction, the normalisation
of ${\cal P}_{data}^{(NN)}({\bm x})$ will depend on the goodness of the Neural Network approximation
to the conditional probability, so that in general it must be normalised a-posteriori. 
In the artificial (High Energy Physics) example shown below, the normalisation of the obtained p.d.f.s was consistent with 1
at the 1\% (3\%) level.

On the other hand, one would like to test the goodness of the approximation of the mapped p.d.f.
to the true one. Given a data sample containing $N_{data}$ events, 
it is possible to perform a test 
of the hypothesis of the data sample under consideration being consistent 
with coming from the mapped p.d.f. For that, one can compute the 
distribution of some test statistics like the log-likelihood function of Eq.(\ref{likeli}), 
which can be obtained by generating Monte Carlo samples containing $N_{data}$ events
generated using the mapped p.d.f.
\begin{eqnarray}
{\cal L} = \log(L) = \sum_{e=1}^{N_{data}} \log({\cal P}_{data}^{(NN)}({\bm x}^{(e)}))
\label{likeli}
\end{eqnarray}

Being ${\cal L}_{data}$ the value of the log-likelihood for the original data sample, 
the confidence level ($CL$) associated to the hypothesis of the data sample coming from the
mapped p.d.f. is given by:
\begin{eqnarray}
CL = \int_{-\infty}^{{\cal L}_{data}} d{\cal L} \: {\cal P}({\cal L})
\label{cl}
\end{eqnarray}
\noindent which in practice can be obtained as the fraction of generated Monte Carlo samples of the data size
having a value of the log-likelihood equal or below the
one for the data sample. If the mapped p.d.f. is a good approximation to
${\cal P}_{data}$, the expected distribution for $CL$ evaluated 
for different data samples should have a flat distribution
as it corresponds to a cumulative distribution.

\section{Artificial example}

In this section we propose a purely artificial example in order to illustrate how a Neural Network
can perform a mapping of a 5-dimensional p.d.f. in an unbinned way from
examples.

In this example our "data" will consist in a sample of 100000 events generated in the cube 
$\left [0,\pi\right ]^5 \in {\bf R}^5$ according to the following p.d.f.:
\begin{eqnarray}
{\cal P}_{data}({\bm x}) = \frac{1}{C} \: (\sin(x_1 + x_2 + x_3) + 1)\: 
\biggl(\frac{\sin(x_4^2 + x_5^2)}{x_4^2 + x_5^2} + 1\biggr),
\end{eqnarray}
\noindent which we want to estimate from the generated events. 
In the above expression, $C$ is a normalisation factor such that ${\cal P}_{data}({\bm x})$ has unit integral.
The above p.d.f. has a rather intrincate structure of maxima and minima in both, the 3-dimensional
space of the first three variables and the 2-dimensional space of the two last variables.

In order to map the above p.d.f., we need to train a Neural Network to disentangle between events
generated according to ${\cal P}_{data}({\bm x})$ and events generated according to any
${\cal P}_{ref}({\bm x})$ non-vanishing in any region where ${\cal P}_{data}({\bm x})$ is different 
from zero. In order to make easier the learning of the existing correlations in the
5-dimensional space, as explained before, ${\cal P}_{ref}({\bm x})$ is chosen as 
the product of good approximations to the 1-dimensional projections of ${\cal P}_{data}({\bm x})$, properly normalised
to have unit integral. 

In the case of data p.d.f., 
it turns out that the 1-dimensional projections of the three first variables are equal and
essentially flat, whereas the 1-dimensional projections for the two last variables
can be parametrised as a 4th degree polinomial ($P_4$). Therefore, we choose as reference p.d.f.:
\begin{eqnarray}
{\cal P}_{ref}({\bm x}) = \frac{1}{C'} \: P_4(x_4)\cdot P_4(x_5)
\end{eqnarray}
\noindent and generate a number of 100000 events according to it. As before, $C'$ is a normalisation factor so that
${\cal P}_{ref}({\bm x})$ has unit integral.

\begin{figure}[p]
\begin{center}
\mbox{
\epsfig{file=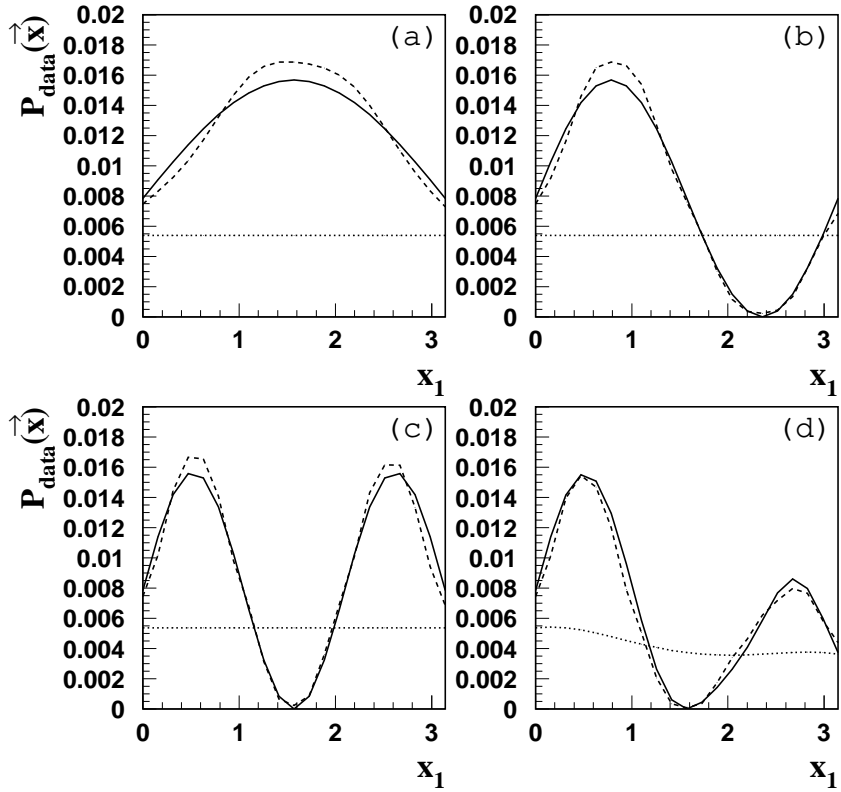,width=12cm}
}
\end{center}
\caption{\protect\footnotesize
Comparison between the true (solid line), the mapped
(dashed line) and the reference (dotted line) p.d.f. versus $x_1$
for different slices in the 5-dimensional space: (a) $x_2 = x_3 = x_4 = x_5 = 0$,
(b) $x_2 = x_1$, $x_3 = x_4 = x_5 = 0$, (c) $x_3 = x_2 = x_1$, $x_4 = x_5 = 0$ and
(d) $x_4 = x_3 = x_2 = x_1$, $x_5 = 0$.}
\label{mathex}
\end{figure}

After the training and normalisation, the p.d.f. given by Eq.(\ref{pdf})
constitutes a reasonably good approximation to ${\cal P}_{data}({\bm x})$, as it is indeed observed in
Fig.~\ref{mathex}, where both are compared for different slices in the 5-dimensional space with
respect to the variable $x_1$. For comparison, it is also shown the reference p.d.f. which,
as expected, is unable to reproduce the complicated structure of maxima and minima in the
5-dimensional space.

\begin{figure}[p]
\begin{center}
\mbox{
\epsfig{file=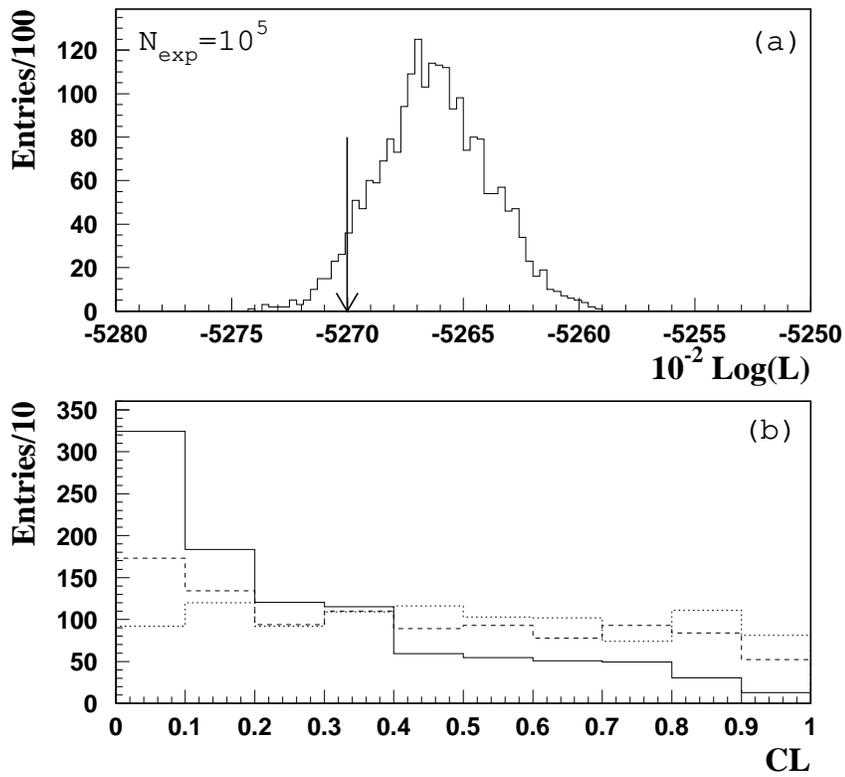,width=12cm}
}
\end{center}
\caption{\protect\footnotesize
(a) Distribution of the log-likelihood computed for Monte Carlo samples
of 100000 events generated according to the mapped p.d.f. The arrow indicates the value of the
log-likelihood for the original data sample.
(b) Distribution of the confidence level for data samples containing
1000 (dotted line), 10000 (dashed line) and 100000 (solid line) events respectively, 
generated with the true p.d.f., of being consistent with the hypothesis of coming from the
mapped p.d.f.}
\label{mathex2}
\end{figure}

As explained in previous section, it is posible to perform a test of 
the goodness of the mapped p.d.f.
For that, a number of 10000 Monte Carlo samples have been generated with the mapped
p.d.f., each one containing 100000 events, which is the same number of events of the
"data" sample. 
The log-likelihood is computed for each MC sample and its
distribution is shown in Fig.~\ref{mathex2}a), in which the arrow indicates the value of the
log-likelihood for the original data sample (${\cal L}_{data}$). From this distribution and the value of ${\cal L}_{data}$ 
we have found a confidence level of 5.5\%
associated to the hypothesis of the data sample coming from the
mapped p.d.f. This seems a low CL  and needs further comments,
but  as we know the true p.d.f given by Eq.(\ref{pdf}), 
we can do much better than performing a single  
measurement for $CL$ and is to find out its distribution.

Very often in High Energy Physics and other fields the problem consist on estimating a p.d.f. from a
sample of simulated Monte Carlo events which is much larger  (typically a factor 100
times larger) than the experimental data sample over which we should use this p.d.f (see the High Energy Physics
example of Sect. 4). For this reason we have obtained the $CL$ distribution in three different scenarios:
when the number of experimental data events ($N_{exp}$ )  has the same number of events 
as the data sample used to obtain the mapped p.d.f.
($N_{data}$ = 100000), and two with smaller statistics, one with $N_{exp}$ = 10000 and another 
with $N_{exp}$ = 1000.

A number of 10000 Monte Carlo samples have been generated with the mapped
p.d.f., each containing $N_{exp}$ events, for the three different values of
$N_{exp}$ and the log-likelihood is computed for each sample in all three scenarios. 
On the other hand, a number of 1000 data samples are generated with the true
p.d.f. in the three scenarios and the confidence level is computed according to Eq.(\ref{cl}). 
The distribution of $CL$ is shown in Fig.~\ref{mathex2}b) for $N_{exp}=1000$ (dotted line),
10000 (dashed line) and 100000 (solid line). 
It can be observed that for $N_{exp}=1000$ the distribution 
of $CL$ is to a good approximation a flat distribution whereas for $N_{exp}=10000$ 
it starts deviating from being flat, which indicates that the statistics of the data sample
is high enough to start ``detecting'' systematic deviations in the mapped p.d.f. with respect to
the true one. 

In the case of $N_{exp}=1000$ which, as mentioned above
illustrates a common situation in High Energy Physics, the mapped p.d.f. 
turns out to be a good enough approximation 
when used for the smaller experimental data sample.
In the other extreme, $N_{exp}=100000$, which illustrates 
the situation in which there is a unique data sample from which
one wants to  estimate the underlying p.d.f.,  it can be observed
in Fig.~\ref{mathex2}b) (solid line) the existence of enough resolution to detect systematic deviations 
in the mapped p.d.f. with respect to the true one. 
It should be stressed the very complicated structure of the true p.d.f., which makes
extremely difficult its accurate mapping and 
nevertheless the difference between both distributions 
are the ones observed in Fig.~\ref{mathex} between  the solid and the dashed lines. 
In such situations we can not use the mapped p.d.f. for fine probability studies but it is clear that it is
still very useful for other kind of studies like classification or discrimination. 

\section{High Energy Physics example}

In order to illustrate the practical interest of p.d.f. mapping, the following 
High Energy Physics example is considered. 

One of the major goals of LEP200 is the precise measurement of the mass of the
W boson. At energies above the WW production threshold ($\sqrt{s} > 161$ GeV)
W bosons are produced in pairs and with sufficient 
boost to allow a competitive measurement of the W mass by direct reconstruction
of its product decays. 
Almost half of the times ($45.6\%$) both W bosons decay hadronically,
so that four jets of particles are observed in the final state. 
 
Most of the information about the W mass is contained in the reconstructed di-jet invariant
mass distribution, so that $M_W$ can be estimated by performing a likelihood fit
to this 2-dimensional distribution.
Therefore, the W mass estimator, $\hat{M}_W$, is obtained by maximising the
log-likelihood function:

\begin{equation}
{\cal L}(M_W) = \sum_{e=1}^{N} \log{\cal P}(s_1'^{(e)},s_2'^{(e)}\mid M_W)
\end{equation}

\noindent with respect to $M_W$, where ${\cal P}(s_1'^{(e)},s_2'^{(e)} \mid M_W)$ represents the
probability of event $e$, characterised by the two measured invariant masses
$(s_1'^{(e)},s_2'^{(e)})$, 
given $M_W$ which, accounting for the existing background, can be expressed as:

\begin{equation}
{\cal P}(s_1',s_2'\mid M_W) = \rho_{ww} {\cal P}_{ww}(s_1',s_2'\mid M_W) +
(1-\rho_{ww}) {\cal P}_{bckg}(s_1',s_2').
\end{equation}

In the above expression $\rho_{ww}$ is the expected signal purity in the sample and
${\cal P}_{ww}$ and ${\cal P}_{bckg}$ 
are respectively the p.d.f. for signal (W-pair production) and background in terms
of the reconstructed di-jet invariant masses. For a typical selection procedure
above threshold at LEP200, signal efficiencies in excess of 80\% with a purity at the
level 80\% can be obtained in the fully hadronic decay channel.

Therefore, in order to determine $M_W$, we need to obtain both
p.d.f.s, for signal and background, in terms of the reconstructed di-jet invariant masses. 

At $\sqrt{s} = 172$ GeV and after selection, most of the background comes from QCD. 
To map the p.d.f. for the background, a 2-5-2-1 Neural Network was trained with $\sim 6000$ 
selected $q\bar{q}$ Monte Carlo events generated with full detector simulation (``data'')
and the same number of ``reference'' 
Monte Carlo events generated according to
the 1-dimensional projections of the ``data'' sample.

As far as the signal p.d.f. is concerned, it depends on the parameter we want to
estimate: $M_W$. It can be obtained by a folding procedure of the theoretical 
prediction for the 3-fold differential cross-section in terms of the 2 di-quark invariant masses
($s_1$ and $s_2$) and $x$ (the fraction of energy radiated in the form of initial state photons),
with a transfer function $T$, which accounts for distortions in the kinematics of the 
signal events due
to fragmentation, detector resolution effects and biases in the reconstruction
procedure. This transfer function represents the conditional 
probability of the reconstructed invariant masses given some invariant masses at the parton level 
and initial state radiation (ISR). The ISR
is most of the times lost along the beam pipe and therefore unknown, reason
for which it must be integrated over. This conditional probability is given by:
\begin{equation}
T(s_1',s_2' \mid s_1,s_2,x) = \frac{f(s_1',s_2',s_1,s_2,x)}
{g(s_1,s_2,x)}, 
\end{equation}
\noindent where $s_i'$ stands for each reconstructed invariant mass and
$g(s_1,s_2,x)$ is theoretically known and
has a compact expression, reason for which there is no need to map it. 

Then, the goal is to map the 5-dimensional p.d.f. $f(s_1',s_2',s_1,s_2,x)$. To do it, 
a 5-11-5-1 Neural Network was trained with 40000 hadronic WW Monte Carlo events
generated with full detector simulation (``data'')
and the same number of ``reference'' events 
generated according to the 1-dimensional projections of the ``data'' sample.

\begin{figure}[p]
\begin{center}
\mbox{
\epsfig{file=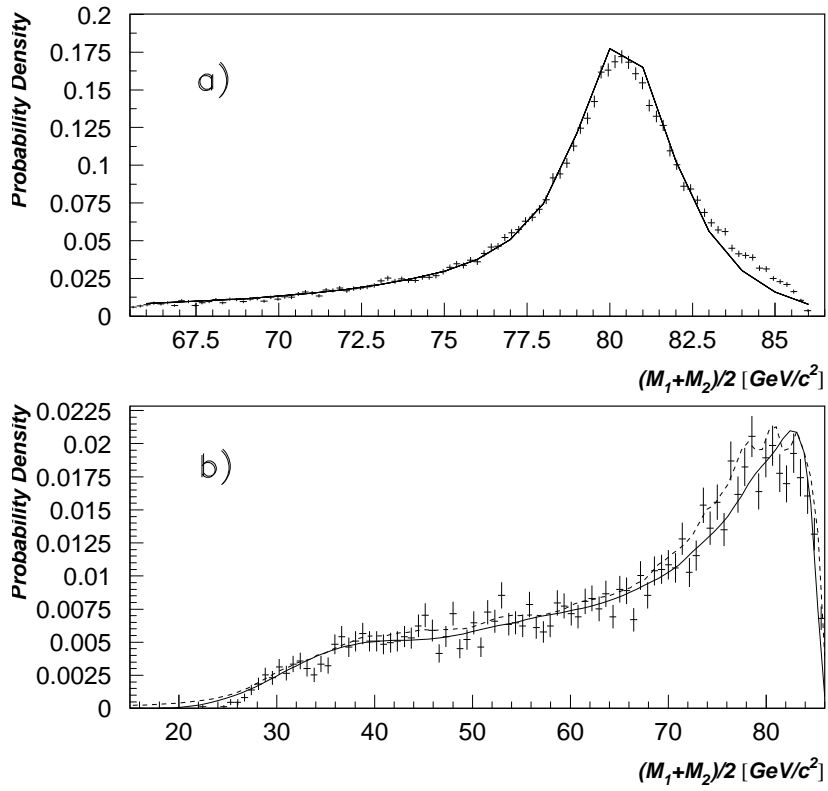,width=12cm}
}
\end{center}
\caption{\protect\footnotesize
Comparison between NN (solid line) and Monte Carlo (points with
error bars) prediction for the
average di-jet invariant mass p.d.f. for a) signal and
b) background. In b), the p.d.f. as obtained by a box method
(dashed line) is also shown.}
\label{nnmapping}
\end{figure}

In order to test that the
event-by-event p.d.f. is meaningful, the predicted 1-dimensional
projection of the average invariant mass distribution is compared to Monte Carlo 
in Figs.~\ref{nnmapping}a) and b) for both signal and background
by using the obtained ${\cal P}_{ww}$ and ${\cal P}_{bckg}$, respectively.
Note the overall good agreement between the distributions.

The unbiasedness of the obtained estimator is checked by computing
the calibration curve with respect the true parameter by performing
a large number of fits to Monte Carlo samples generated with different values
of $M_W$.

The performance of the NN in mapping a n-dimensional p.d.f.
has been compared to the ``box method''~\cite{Schmidt}, a standard procedure
to build up binned p.d.f.s. 
In the case of the background p.d.f., which is only 2-dimensional, the ``box method'' 
yielded
reasonable results as shown in Fig.~\ref{nnmapping}b), while
in the case of the 5-dimensional p.d.f.
it showed strong limitations which made impossible its application.
The main reason is the time required to compute the final p.d.f which needs an integration
on top of the adjustement of the ``box method'' parameters
(initial box size, minimum number of MC points inside each box, etc)
in a space of high dimensionality and limited statistics.
Is in this environment where the mapping of p.d.f.s by means of NNs may be superior
to ``standard binned procedures'' in terms of accuracy (the p.d.f. is determined in an
unbinned way from examples) and speed (the resulting p.d.f. is an analytic function).

\section{Conclusions}

We have shown that Neural Networks
are useful tools for building up $n$-di\-men\-sio\-nal 
p.d.f.s from examples in an unbinned way. The method takes
advantage of the interpretation of the Neural Network output, after training, in terms of 
a-posteriori Bayesian probability when a unary representation is taken for the output patterns.
A purely artificial example and
an example from High Energy Physics, in which the mapped p.d.f.s are used to determine
a parameter through a maximum likelihood fit, have also been discussed.
In a situation of high dimensionality of the space to be mapped and limited available
statistics, the method is superior to ``standard binned procedures''.

\section{Acknowledgements}

This research has been partly supported by CICYT under contract number
AEN97-1697.

\newpage

\section*{Figure captions}

\begin{itemize}
\item {\bf Figure 1:}
Comparison between the true (solid line) and the mapped
(dashed line) and the reference (dotted line) p.d.f. versus $x_1$
for different slices in the 5-dimensional space: (a) $x_2 = x_3 = x_4 = x_5 = 0$,
(b) $x_2 = x_1$, $x_3 = x_4 = x_5 = 0$, (c) $x_3 = x_2 = x_1$, $x_4 = x_5 = 0$ and
(d) $x_4 = x_3 = x_2 = x_1$, $x_5 = 0$.

\item {\bf Figure 2:}
(a) Distribution of the log-likelihood computed for Monte Carlo samples
of 100000 events generated according to the mapped p.d.f. The arrow indicates the value of the
log-likelihood for the original data sample.
(b) Distribution of the confidence level for data samples containing
1000 (dotted line), 10000 (dashed line) and 100000 (solid line) events respectively, 
generated with the true p.d.f., of being consistent with the hypothesis of coming from the
mapped p.d.f.

\item {\bf Figure 3:}
Comparison between NN (solid line) and Monte Carlo (points with
error bars) prediction for the
average di-jet invariant mass p.d.f. for a) signal and
b) background. In b), the p.d.f. as obtained by a box method
(dashed line) is also shown.

\end{itemize}

\newpage

\newpage

\newpage

\end{document}